\def\twocol{2}

\def\colopt{\twocol}
\ifnum\colopt=\twocol
 \documentstyle[floats,twocolumn,aps,psfig]{revtex}
 \def\Section#1{}
\else
 \documentstyle[floats,preprint,aps,psfig]{revtex}
 \def\Section#1{\section{#1}}
\fi

\def\Sv{\vec{S}}
\def\Sz{S^z}
\def\Jp{{J'}}
\def\abs#1{\vert #1 \vert}
\def\beq{\begin{equation}}
\def\eeq{\end{equation}}
\def\bea{\begin{eqnarray}}
\def\eea{\end{eqnarray}}
\def\nn{\nonumber}
\def\ie{{\it i.e.}}
\def\Pdiag{\phi_{{\rm diag}}}
\def\HAF{HAF}	
\def\PBC{PBC}	
\def\OBC{OBC}	
\def\MPD{Magnetic phase diagram}	
\def\mPD{magnetic phase diagram}	
\font\amsmath=msbm10

\def\Zed{\hbox{\amsmath Z}}

\begin{document}
\tolerance 50000
\preprint{
\begin{minipage}[t]{1.8in}
\rightline{La Plata-Th 97/12}
\rightline{SISSA 88/97/EP}
\rightline{}
\end{minipage}
}

\draft

\ifnum\colopt=\twocol
 \twocolumn[\hsize\textwidth\columnwidth\hsize\csname @twocolumnfalse\endcsname
\fi

\title{Magnetization Curves of Antiferromagnetic
       Heisenberg Spin-${1 \over 2}$ Ladders}

\author{D.C.\ Cabra$^{1}$, A.\ Honecker$^{2,\dag}$, P.\ Pujol$^{2}$
}
\address{
$^{1}$Departamento de F\'{\i}sica, Universidad Nacional de la Plata,
      C.C.\ 67, (1900) La Plata, Argentina.\\
$^{2}$International School for Advanced Studies,
      Via Beirut 2-4, 34014 Trieste, Italy.\\
$^{\dag}$Work done under support of the EC TMR Programme
         {\em Integrability, non-per\-turba\-tive effects and symmetry in
         Quantum Field Theories}, grant FMRX-CT96-0012.\\
}

\date{July 9, 1997, revised October 13, 1997}
\maketitle
\begin{abstract}
\begin{center}
\parbox{14cm}{
Magnetization processes of spin-${1 \over 2}$ Heisenberg ladders
are studied using strong-coupling expansions, numerical diagonalization
of finite systems and a bosonization approach. We find that the
magnetization exhibits plateaux as a function of the applied field
at certain rational fractions of the saturation value. Our main
focus are ladders with 3 legs where plateaux with magnetization
one third of the saturation value are shown to exist.
}
\end{center}
\end{abstract}

\ifnum\colopt=\twocol
 \vskip1pc]
\fi

\Section{Introduction}

Recently there has been considerable interest in
coupled Heisenberg antiferromagnetic (\HAF) chains,
so-called `spin ladders', where one of the fascinating
discoveries is that the appearance of a gap depends on
the number of chains being even or odd (for recent reviews
see e.g.\ \cite{review}). 
In this letter we study ladder systems at zero temperature in
a strong uniform magnetic field. This issue has so far only
been addressed for {\it two} coupled chains with a magnetization
experiment on Cu$_2$(C$_5$H$_{12}$N$_2$)$_2$Cl$_4$ \cite{CCLMMP}
and theoretically using numerical diagonalization \cite{HLP},
series expansions \cite{WOS} and a bosonization approach \cite{ChiGi}.
These studies found a plateau at zero magnetization whose width is
given by the spin gap in the otherwise smooth magnetization curve.
In this letter we extend the theoretical approaches to
three and more coupled chains using strong coupling expansions,
numerical diagonalization and a bosonization approach. We find
that in general the magnetization curves exhibit plateaux also 
at certain non-zero quantized values of the magnetization.

To be precise, we concentrate on the zero-tempera\-ture
behaviour of the following {\HAF} 
spin ladder with $N$ legs (kept fixed) and length $L$
(taken to infinity):
\bea
H^{(N)} &=& \Jp \sum_{i} \sum_{x=1}^L \Sv_{i,x} \Sv_{i+1,x}
     + J \sum_{i=1}^N \sum_{x=1}^L \Sv_{i,x} \Sv_{i,x+1} \nn \\
    && - h \sum_{i,x} \Sz_{i,x} \, ,
\label{hamOp}
\eea
where the $\Sv_{i,x}$ are spin-${1 \over 2}$ operators and
$h$ is a dimensionless magnetic field. We assume
periodic boundary conditions along the chains but investigate
both open (\OBC) and periodic boundary conditions (\PBC) along the rungs.
The magnetization $\langle M \rangle$ is given by the
expectation value of the operator
$M = {2 \over L N} \sum_{i,x} \Sz_{i,x}$.

Our main result is that $\langle M \rangle$ as a function of $h$ has plateaux
at the quantized values
\beq
{N \over 2} (1 - \langle M \rangle) \in \Zed \, .
\label{condM}
\eeq
This condition also appears in the Lieb-Schultz-Mattis theorem \cite{LSM}
and its generalizations \cite{Affleck,AOY}. There, it is related to gapless
{\it non-magnetic} excitations, but plateaux in magnetization curves appear
if there is a gap to {\it magnetic} excitations.

\Section{Strong coupling limit}

However, there are other arguments which lead to (\ref{condM})
as the quantization condition for $\langle M \rangle$
at a plateau. A particularly simple one is given by
the limit of strong coupling along the rungs $\Jp \gg J$
(which has also proven useful in other respects \cite{BDRS}).
At $J = 0$ one has to deal only with Heisenberg chains of
length $N$ and the only possible values for the magnetization
are precisely the solutions of (\ref{condM}):
$\langle M \rangle \in \{- 1, -1 + 2/N, \ldots, 1 - 2/N, 1\}$.

For $N=2$ and $\Jp \gg J$ this consideration predicts a
plateau at $\langle M \rangle = 0$. The boundary of this
plateau is related to the spin gap simply by $h_c = \Delta$.
The strong-coupling series for this gap reads \cite{RRT}
\beq
\Delta = \Jp - J + {J^2 \over 2 \Jp}
+ {J^3 \over 4 \Jp^2} - {J^4 \over 8 \Jp^3} + {\cal O}(J^5) \, ,
\label{Delta2}
\eeq
where we have extended the result of \cite{RRT} to fourth order
(this further order is crucial to obtain a zero of the
gap for $\Jp \ge 0$).

For $N=3$ and small magnetic fields one has a degeneracy which
makes already first-order perturbation theory non-trivial.
For {\OBC} and $\abs{\langle M \rangle} \le 1/3$, the low-lying
spectrum is then given by a spin-${1 \over 2}$ chain in a
magnetic field whose magnetization curve is well-studied
(see \cite{BoFiPa} and references therein). On the other hand,
for {\PBC} and $\abs{\langle M \rangle} \le 1/3$, the first-order
low-energy effective Hamiltonian for (\ref{hamOp}) at $N=3$
turns out to be (see also \cite{ToSuSchulz}):
\bea
H^{(III,p)}_{{\rm eff.}} &=& {J \over 3}
\sum_{x=1}^L \left( 1 + \sigma_x^{+} \sigma_{x+1}^{-}
 + \sigma_x^{-} \sigma_{x+1}^{+} \right) \Sv_x \Sv_{x+1} \nn \\
 && - h \sum_{x=1}^L \Sz_{x} \, ,
\label{lowEn3p}
\eea
where the $\Sv_x$ are $su(2)$ operators acting in the spin-space
and $\sigma_x^{\pm}$ act on another two-dimensional space which
comes from a degeneracy due to the permutational symmetry of the
chains. In particular, the usual spin-rotation symmetry is enlarged
by an $U(1)$ coming from the XY-type interaction in the space
of the Fourier transforms along the rungs (first factor in
(\ref{lowEn3p})). The Hamiltonian (\ref{lowEn3p})
encodes the magnetization curve in the strong-coupling limit
of the three-leg ladder with {\PBC} for
$\abs{\langle M \rangle} \le 1/3$.

At least for {\OBC}, the situation is a little more favourable for
the $\langle M \rangle = 1/3$ plateau that one expects for $N=3$. 
It turns out that one can use non-degenerate perturbation theory
to compute the energy cost to flip a spin around this plateau and
thus its lower and upper boundaries:
\bea
h_{c_1} &=& 2 J - {2 J^2 \over 9 \Jp} - {94 J^3 \over 243 \Jp^2}
 + {\cal O}(J^4) \, ,
\label{hc1} \\
h_{c_2} &=& {3 \over 2} \Jp - J
+ {3 J^2 \over 4 \Jp} + {2065 J^3 \over 3888 \Jp^2} + {\cal O}(J^4) \, .
\label{hc2}
\eea

Finally, it is straightforward to exactly compute the upper critical
field $h_{uc}$ for the transition 
to a fully magnetized state for {\PBC}, even $L$ and arbitrary $N$; 
\beq
h_{uc} = \cases{2 \left(J + \Jp \right) & $N$ even, \cr
     \left(1-\cos\left(\pi {N-1 \over N}\right)\right) \Jp + 2 J
                                           & $N$ odd. \cr}
\label{maxTrans}
\eeq
Actually, the result (\ref{maxTrans}) at $N=3$,
$h_{uc} = 3 / 2 \Jp + 2 J$ also applies to {\OBC}.

\Section{Numerical results}

We now proceed with considerations that follow closely classical
work on single Heisenberg spin chains \cite{BoFiPa} and
the 2D triangular {\HAF} \cite{MiNi}.
We have numerically calculated the
lowest eigenvalues as a function of the magnetization, wave
vectors and coupling constants on finite systems with up to a total
of 24 sites.

\begin{figure}[ht]
\psfig{figure=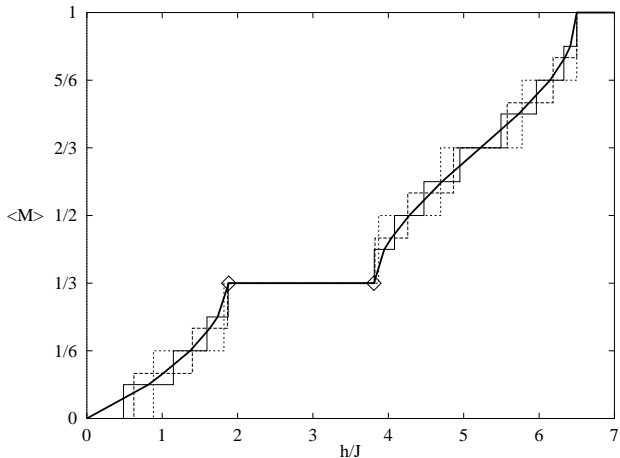,width=\columnwidth,angle=270}
\smallskip
\caption{
Magnetization curve for $N=3$ at $\Jp/J = 3$ with {\OBC}.
The thin full lines are for $L=8$, the long dashed lines for
$L=6$ and the short dashed lines for $L=4$. The thick full
line indicates the expected form in the thermodynamic limit.
The two diamonds denote the series (\ref{hc1}) and (\ref{hc2})
for the boundaries of the plateau.
\label{figure0}
}
\end{figure}

Fig.\ \ref{figure0} illustrates our main results with a
magnetization curve for $N=3$ at $\Jp/J = 3$ with {\OBC}.
The thin lines denote curves at finite system size and
clearly exhibit a plateau with $\langle M \rangle = 1/3$.
The strong-coupling approximations (\ref{hc1}) and (\ref{hc2})
for the boundaries of this plateau are in good agreement
with the finite-size data. Finite-size effects are also
small for the midpoints of the steps in the magnetization
curves \cite{BoFiPa}. From this one obtains an
extrapolation to infinite $L$ which is shown by the
full line in Fig.\ \ref{figure0}.

\begin{figure}[ht]
\psfig{figure=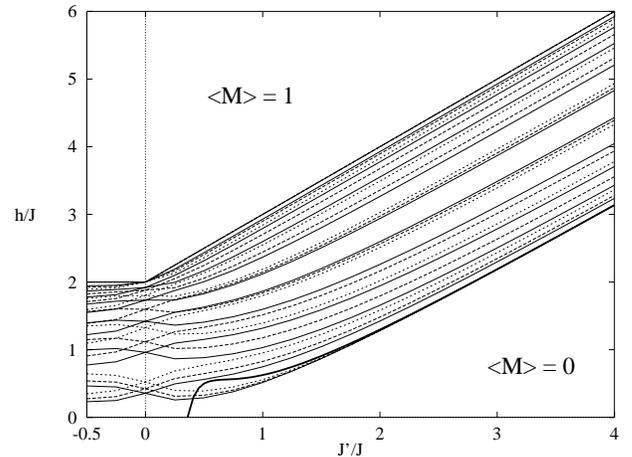,width=\columnwidth,angle=270}
\smallskip
\caption{
{\MPD} of the ladder with $N=2$ legs. The thin full lines
are for $L=12$, the long dashed lines for $L=10$ and the
short dashed lines for $L=8$. The thick full
line shows the gap (\ref{Delta2}).
\label{figure1}
}
\end{figure}

A more compact representation is given by `{\mPD}s' which show
the projection of the conventional magnetization curves as in
Fig.\ \ref{figure0} onto the axis of the magnetic field.
We illustrate this in Fig.\ \ref{figure1} with the case
of a two-leg ladder which has already been studied in \cite{HLP}.
On a finite lattice the possible
values of $\langle M \rangle$ are quantized; the values of the magnetic
field $h/J$ where a transition from one such value occurs to
the neighbouring one are shown by lines as a function of $\Jp/J$. 
Regions without such lines denote plateaux in the magnetization
curve while regions where they are close to each other correspond to
smooth transitions in the thermodynamic limit.

The magnetization curves of \cite{CCLMMP,HLP} with $\Jp/J = 5$
are located somewhat beyond the right border of Fig.\ \ref{figure1}.
Our figure contains additional information about the dependence
on $\Jp/J$.
Note that the numerically obtained values to full magnetization
do indeed match with the version of (\ref{maxTrans}) for the
ladder: $h_{uc} = 2 J + \Jp$. For the boundary of the plateau
at $\langle M \rangle = 0$ one observes that finite-size
effects are not substantial for $\Jp/J \ge 3/2$, and in this
region one observes also good agreement with the approximation
(\ref{Delta2}) for the transition value of the magnetic field
$h_c = \Delta$. At weak coupling,
field theoretic arguments \cite{ToSuLShNeTs} predict the
gap to open linearly, $\Delta \sim \Jp$. This
is compatible with Fig.\ \ref{figure1}, though
due to the large finite-size effects in this region much
longer chains are needed to confirm this linear behaviour
conclusively \cite{GBW}.

\begin{figure}[ht]
\psfig{figure=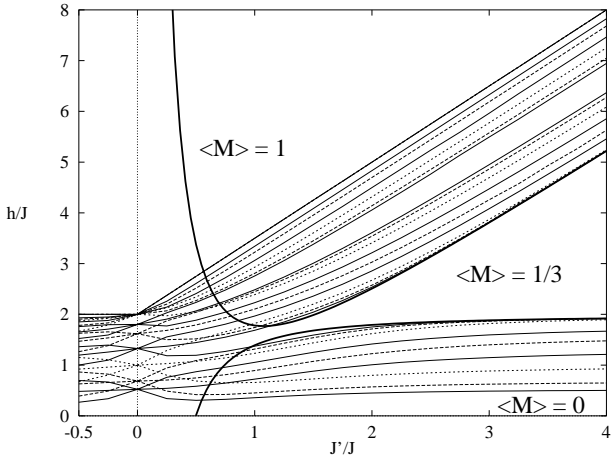,width=\columnwidth,angle=270}
\smallskip
\caption{
{\MPD} of $N=3$ chains with {\OBC}. The thin full lines
are for $L=8$, the long dashed lines for $L=6$ and the
short dashed lines for $L=4$. The two thick full
lines denote (\ref{hc1}) and (\ref{hc2}), respectively.
\label{figure2}
}
\end{figure}

\begin{figure}[ht]
\psfig{figure=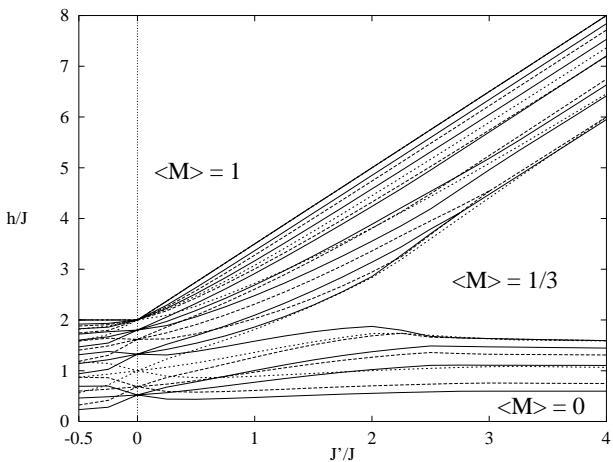,width=\columnwidth,angle=270}
\smallskip
\caption{
Same as Fig.\ \ref{figure2}, but with {\PBC}.
\label{figure3}
}
\end{figure}

Figs.\ \ref{figure2} and \ref{figure3} show the {\mPD}s 
for $N=3$ with {\OBC} and {\PBC}, respectively (Fig.\
\ref{figure0} is a section of Fig.\ \ref{figure2} at
$\Jp/J = 3$). Both figures clearly exhibit a plateau with
$\langle M \rangle = 1/3$ at least in the region $\Jp \ge 2 J$.
The strong-coupling series (\ref{hc1}) and (\ref{hc2}) for the
boundaries of this plateau are also shown in Fig.\ \ref{figure2}
and in the aforementioned region $\Jp \ge 2 J$, where finite-size
effects are small, one observes good agreement between the
expansions and the numerical results.

Figs.\ \ref{figure2} and \ref{figure3} differ at least in their
details. For example, one observes that at
the upper boundary of the $\langle M \rangle = 1/3$
plateau in Fig.\ \ref{figure3} it becomes favourable
to flip two spins at once rather than one for $\Jp \ge 3 J$
which may be interpreted as one signal of the frustration
introduced into the system.

At strong coupling the excitations above the 1/3-plateau in
Fig.\ \ref{figure3} are described by (\ref{lowEn3p}) with all
spins aligned along the field ($S^z_x = {1 \over 2}$).
This is an XY-chain and therefore massless, providing us with
an example of a plateau in the magnetization curve with gapless
non-magnetic excitations above it.

\Section{Abelian bosonization}

The strong-coupling expansions as well as the numerical results
obtained so far clearly show the existence of plateaux for
sufficiently strong coupling. To see if the plateau persists
in the weak-coupling region $\Jp \ll J$
we use abelian bosonization (see e.g.\ \cite{SchBo,AffLe}). 
The computation to be presented
below is similar to the ones performed recently in \cite{AOY,Totsuka}, so
we refer the interested reader to these two references for more details.

The starting point for this weak-coupling expansion is the observation
that at $\Jp = 0$ one has $N$ decoupled spin-${1 \over 2}$
{\HAF}s in a magnetic field whose low-energy
properties are described by a $c=1$ Gaussian conformal field theory
\cite{WET}. This theory is characterized by the radius of
compactification $R(\langle M \rangle, \Delta)$ which depends
on the magnetization as well as the XXZ-anisotropy $\Delta$ which
we have here introduced for convenience. At zero magnetic field one
has \cite{AffLe}: $R^2(0,\Delta) = {1 \over 2 \pi} \left(1 -
{1 \over \pi} \cos^{-1} \Delta\right)$.
In general, this radius can be computed using
the Bethe-Ansatz solution \cite{WET} for the Heisenberg chain.
For a qualitative understanding it may be helpful to notice
that the radius of compactification $R$ decreases if either
$\Delta$ becomes smaller or if the magnetization $\langle M \rangle$
increases.

Now the low-energy properties of the Hamiltonian (\ref{hamOp})
at small coupling are described by the following Tomonaga Hamiltonian
with interaction terms:
\bea
\bar{H}^{(N)} &=& \int {\rm d}x \Biggl[
      {\pi \over 2} \sum_{i=1}^N \left\{ \Pi_i^2(x) +
    R^2(\langle M \rangle, \Delta) \left(\partial_x \phi_i(x)\right)^2 \right\}
\nn \\
& +& {\lambda_1 \over 2 \pi} \sum_i \left(\partial_x \phi_i(x)\right)
      \left(\partial_x \phi_{i+1}(x)\right) \nn \\
& +& \sum_i \biggl\{
            \lambda_2 : \cos\left(4 k_F x + \sqrt{4\pi}(\phi_i + \phi_{i+1})\right) :
\label{LeH} \\
&& \qquad + \lambda_3 : \cos\left( \sqrt{4 \pi} (\phi_i - \phi_{i+1})\right) : \nn \\
&& \qquad + \lambda_4 : \cos\left(\sqrt{\pi}(\tilde{\phi}_i - \tilde{\phi}_{i+1}) \right) :
    \biggr\}\Biggr] \, , \nn
\eea
with $\Pi_i = {1 \over \pi} \partial_x \tilde{\phi}_i$, 
and $\lambda_i \sim \Jp/J$.
It is convenient to rescale the fields by the radius of compactification
$R$ in (\ref{LeH}), \ie\ $\phi_i \to \sqrt{4\pi} R \phi_i$.

For $N=3$ chains with {\PBC} we now change
variables from the fields $\phi_1$, $\phi_2$, $\phi_3$ to
${1 \over \sqrt{3}} \left(\phi_1+\phi_2+\phi_3\right)$,
${1 \over \sqrt{2}} \left(\phi_1-\phi_2\right)$,
${1 \over \sqrt{6}} \left(\phi_1+\phi_2-2\phi_3\right)$.
Following \cite{SchBo} one can show that the perturbation terms
with coefficients $\lambda_3$ or $\lambda_4$ in (\ref{LeH}) give
rise to a mass for the latter two fields.

Let us now consider the remaining field $\Pdiag := {1 \over \sqrt{N}}
\sum_{i=1}^N \phi_i$. Radiative corrections to (\ref{LeH}) generate
the interaction term $:\cos(2 \Pdiag/R):$ \cite{footnote}.
For $N=3$ and at $\langle M \rangle = 0$
this operator is irrelevant (in the region of $\Delta$ close to 1), which
confirms that three chains weakly coupled in a periodic manner are
massless.

Now we address the question of the appearance of plateaux
for $\langle M \rangle \ne 0$. Since this requires a gap for the
(magnetic) excitations, we expect such a plateau to occur if
the remaining field $\Pdiag$ acquires a mass. In fact, the
additional interaction term
\beq
: \cos(\Pdiag/R) :
\label{cosPdiag}
\eeq
survives in the continuum limit on a plateau \cite{CHPprep}.
The operator (\ref{cosPdiag}) can appear only if (\ref{condM}) holds,
as can easily been inferred from translational invariance of
the original Hamiltonian (\ref{hamOp}) and the fact that
$k_F = {1 \over 2}(1 - \langle M \rangle ) \pi$ and a one-site
translation of the lattice Hamiltonian (\ref{hamOp}) translates
into the internal symmetry transformation $\Pdiag \to \Pdiag + 2 N R k_F$.

For $N=3$, (\ref{condM}) requires that $\langle M \rangle = 1 / 3$.
If one now estimates the radius of compactification $R({1 \over 3},1)$
following e.g.\ \cite{Totsuka}, one finds that at $\Jp = 0$ the operator
(\ref{cosPdiag}) is slightly irrelevant for $\Delta = 1$. The
dimension of this operator decreases with $\Jp$ implying that the
$1/3$-plateau in Fig.\ \ref{figure3} extends
down into the region of small $\Jp$. It should be noted that the
appearance of the plateau for a given small $\Jp$ crucially
depends on the value of $\Delta$, explaining why the numerical
evidence in Fig.\ \ref{figure3} is not conclusive in this region.
For simplicity we have concentrated on {\PBC}. The case of {\OBC}
is qualitatively similar but more subtle in the details and will be
discussed in \cite{CHPprep}.

\Section{Conclusions}

In this letter we have shown that spin ladders exhibit plateaux
in their magnetization curves when subjected to strong magnetic
fields. We have mainly concentrated on the plateau with
$\langle M \rangle = 1/3$ in three coupled chains, but also
other rational values can be obtained by varying the number
of chains $N$. Quantum fluctuations (\ie\ the choice of the spin
$S={1 \over 2}$) do play an important r\^ole --
the plateaux disappear in general when
one inserts classical spins into (\ref{hamOp}). Nevertheless, by
analogy to the two-dimensional triangular antiferromagnet \cite{ChuGo}
it seems likely that such plateaux would appear in first-order
spin-wave theory. Also with Ising spins the behaviour of (\ref{hamOp})
is different: At zero temperature the magnetization changes only
discontinuously between plateaux values (\ie\ no smooth transitions occur),
and a gap opens both for odd and even numbers of chains.

It would be highly interesting to check experimentally whether
the plateaux can indeed be observed, in particular
since nowadays materials with a given number of legs can be
engineered. The first non-trivial check would be to look for
the $\langle M \rangle = 1/3$ plateau in a three-leg ladder.
Here, the material Sr$_2$Cu$_3$O$_5$ \cite{SrCuO} comes to
mind, which is however not very well suited for these purposes
since the necessary order of magnetic fields is not accessible
today due to its large coupling constants. However, there are at
least two-leg ladder materials such as the conventional (VO)$_2$P$_2$O$_7$
\cite{VOPO} or Cu$_2$(C$_5$H$_{12}$N$_2$)$_2$Cl$_4$ \cite{CCLMMP}
with much weaker coupling constants. A three-leg analogue of
such materials could provide a testing ground for our predictions,
in particular if such a material can be found with $\Jp \ge 2 J$
where we would expect a clearly visible plateau in the magnetization
curve at sufficiently low temperatures (cf.\ Fig.\ \ref{figure0}).

\smallskip

A.H.\ thanks Profs.\ K.D.\ and U.\ Schotte for drawing
his attention to magnetization curves. We are grateful to them,
M.\ Kaulke, G.\ Mussardo, A.A.\ Nersesyan, V.\ Rittenberg and
F.A.\ Schaposnik for useful discussions and comments on the manuscript
as well as to G.\ Chaboussant and L.P.\ L\'evy for explanations 
concerning the experimental possibilities.
D.C.C.\ thanks CONICET and Fundaci\'on Antorchas for financial support.
We are indebted to the Max-Planck-Institut f\"ur Mathematik, Bonn-Beuel
for allocation of CPU time on their computers.


\begin{references}
\bibitem{footnote}The appearance of the factor 2 here is intimately
              linked to the fact that we are considering an odd number of
              chains.
\end{references}
\end{document}